\documentclass [10pt, twocolumn, conference] {IEEEtran}
\usepackage[perpage,symbol]{footmisc}
\usepackage {times}
\usepackage {mathptmx, array, paralist}
\usepackage {amsmath}
\usepackage[english]{babel}
\usepackage {graphicx,color}
\usepackage{caption}
\usepackage{subcaption}
\usepackage{epstopdf}
\usepackage{algorithm,algpseudocode}
\usepackage{amsmath,amssymb,makeidx,cite,enumerate}
\usepackage{array}
\usepackage{multirow}
\usepackage{fixltx2e} %For subscript in normal mode
\usepackage{hyphenat}
\usepackage{cite}
\usepackage[numbers,sort&compress]{natbib}
\usepackage{indentfirst}

\usepackage{tabularx}
\usepackage{etoolbox}
\usepackage{verbatim}
\usepackage{color, soul}
\usepackage{enumitem}

\graphicspath {{./figures}}

\begin{document}

\thispagestyle{plain}
\pagestyle{plain}
\pagenumbering{gobble}

\title{RTL-PSC: Automated Power Side-Channel Leakage Assessment at Register-Transfer Level}

\author{\IEEEauthorblockN{Miao (Tony) He\textsuperscript{1}, Jungmin Park\textsuperscript{1}, Adib Nahiyan\textsuperscript{1}, Apostol Vassilev\textsuperscript{2}, Yier Jin\textsuperscript{1}, and Mark Tehranipoor\textsuperscript{1}}
\IEEEauthorblockA{Department of Electrical and Computer Engineering, University of Florida, Gainesville, FL 32611\textsuperscript{1}\\
National Institute of Standards and Technology, Gaithersburg, MD 20899\textsuperscript{2}\\
Email: \{tonyhe, jungminpark, adib1991\}ufl.edu, apostol.vassilev@nist.gov, \{yier.jin, tehranipoor\}@ece.ufl.edu}
\\[-5mm]}

%\IEEEaftertitletext{\vspace{-3in}}

\maketitle

\setlength\parindent{1em}

\begin{abstract}
%\boldmath

Power side-channel attacks (SCAs) have become a major concern to the security community due to their non-invasive feature, low-cost, and effectiveness in extracting secret information from hardware implementation of cryto algorithms. Therefore, it is imperative to evaluate if the hardware is vulnerable to SCAs during its design and validation stages. Currently, however, there is little known effort in evaluating the vulnerability of a hardware to SCAs at early design stage. In this paper, we propose, for the first time, an automated framework, named RTL-PSC, for power side-channel leakage assessment of hardware crypto designs at register-transfer level (RTL) with built-in evaluation metrics. RTL-PSC first estimates power profile of a hardware design using functional simulation at RTL. Then it utilizes the evaluation metrics, comprising of KL divergence metric and the success rate (SR) metric based on maximum likelihood estimation to perform power side-channel leakage (PSC) vulnerability assessment at RTL. We analyze Galois-Field (GF) and Look-up Table (LUT) based AES designs using RTL-PSC and validate its effectiveness and accuracy through both gate-level simulation and FPGA results. RTL-PSC is also capable of identifying blocks\footnote{`design' refers to top module as well as its constituting sub-modules, while a `block' refers to a sub-module within the design.} inside the design that contribute the most to the PSC vulnerability which can be used for efficient countermeasure implementation.

\end{abstract}


\section{Introduction}

Power side-channel attacks (SCAs) exploit the weaknesses in the hardware implementations of crypto algorithms to leak sensitive information, e.g., the encryption key, irrespective of the mathematical robustness of the algorithms. A number of SCAs namely Simple Power Analysis \cite{Kocher:1999:DPA:646764.703989}, Differential Power Analysis (DPA) \cite{Kocher:1999:DPA:646764.703989}, Correlation Power Analysis (CPA) \cite{10.1007/978-3-540-28632-5_2}, Template Attacks \cite{10.1007/3-540-36400-5_3}, Mutual Information Analysis (MIA) \cite{10.1007/978-3-540-85053-3_27}, and Partitioning Power Analysis (PPA) \cite{10.1007/11894063_14} have been proposed and successfully demonstrated over the past two decades. These attacks exploit the fact that the power consumption of a cryptographic hardware depends on the data it processes and the operation it performs \cite{Mangard:2007:PAA:1208234}. 

To counter these attacks, a variety of countermeasures have been proposed to make the crypto hardware resilient to SCAs. The goal of these countermeasures is to reduce the dependencies between the intermediate values of the cryptographic algorithms and the power consumption of the cryptographic devices. For example, hiding countermeasures attempt to break the link between the processed data values and the power consumption of the devices \cite{Mangard:2007:PAA:1208234}. However, all of the SCA countermeasures adversly affect the circuit area, thus making them impractical to be applied to modern resource-constrained designs. One major reason may come from the fact that the evaluation methodology for side-channel leakage assessment are not capable of identifying the source of vulnerabilities effectively and accurately. Hence, the corresponding countermeasure has to be applied to the entire design instead of specific sub-blocks responsible for power side-channel leakage (PSC) vulnerability.

Apart from power side-channel attacks and their corresponding countermeasures, another highly important topic in this domain is power side-channel leakage assessment. Several techniques have been proposed in this domain including signal-to-noise ratio (SNR) \cite{Messerges:2002:ESS:570513.570522}, \iffalse$t$-statistic (i.e., Test Vector Leakage Assessment (TVLA)) \cite{Goodwill11p.:a}\fi Test Vector Leakage Assessment (TVLA) methodology \cite{becker2013test}, success rate \cite{10.1007/11894063_2}, and autonomous side-channel vulnerability evaluator (AMASIVE) \cite{DBLP:conf/birthday/HussSZ13}, \cite{jlpea7010004}. However, these techniques suffer from the following major limitations. 

\begin{itemize}[leftmargin=*]
    \item   They mostly focus on the post-silicon side-channel assessment, which is too late and prohibitively expensive in making any changes to the design to address the leakage issue. 
    \item   Existing techniques typically require large amount of plaintexts and power traces, hence need prohibitively large simulation time, i.e., these techniques are feasible for the post-silicon stage rather than the pre-silicon stage.
    \item   Some techniques, e.g., TVLA \cite{becker2013test} and $\chi^2$-test \cite{moradi2018leakage} can only provide a pass/fail test and cannot give quantitative measure of PSC vulnerability which can lead to false positive results.
    \item   Existing techniques mostly require a security analysts to be manually involved in leakage assessment, which may not be feasible due to cost and time-to-market constraints for modern devices.
\end{itemize}

\begin{figure}[!htbp]
   \vspace{-2mm}
   \centering
   \includegraphics[width=0.4\textwidth]{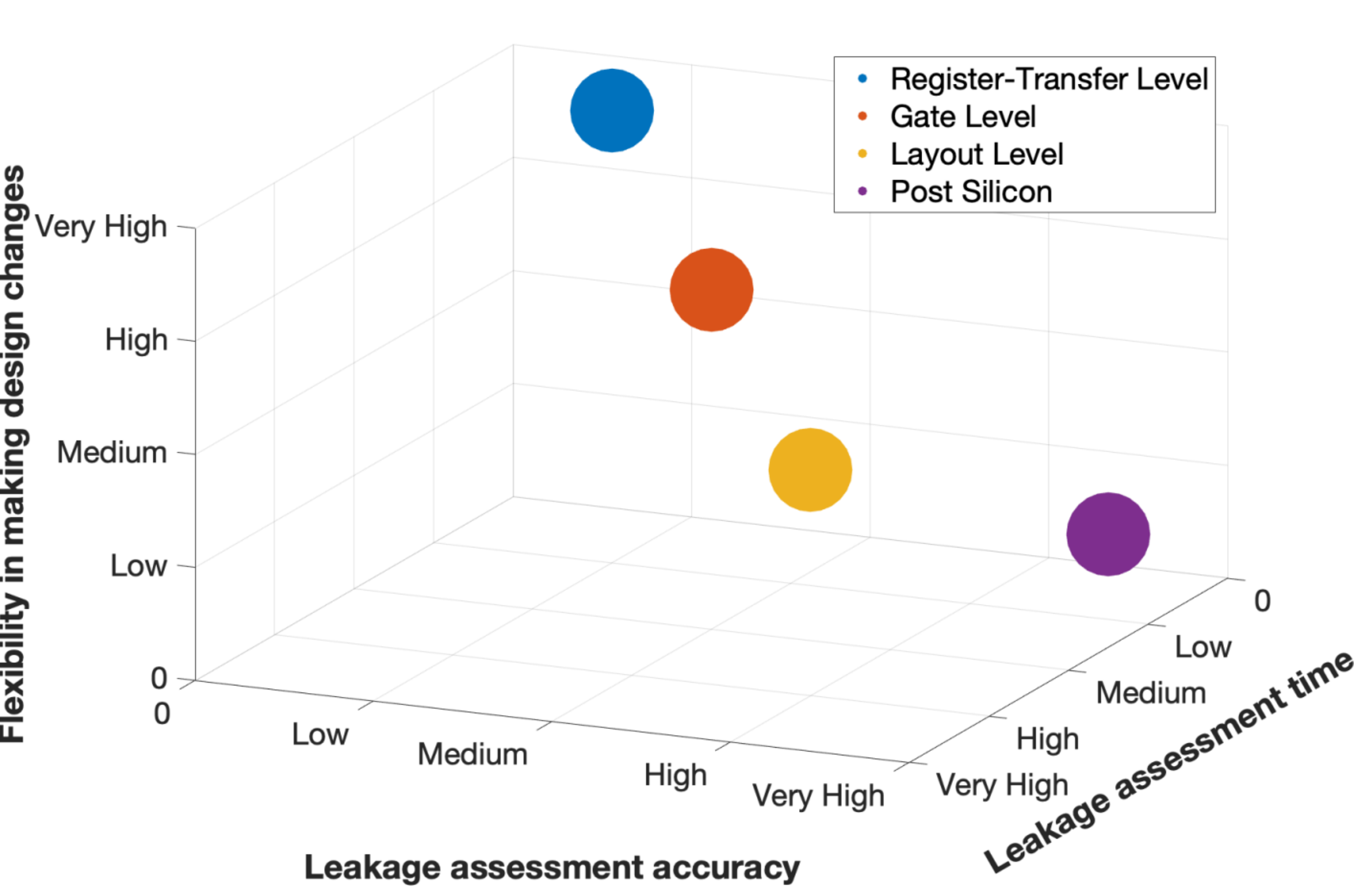}
   \caption{Comparison of the leakage assessment at various stages of the design process.}
   \label{fig:Comparison of the leakage assessment at pre-silicon stage}
  % \vspace{-2mm}
\end{figure}

We summarize the evaluation time and accuracy of side-channel leakage assessment as well as the flexibility to make design changes at different pre-silicon design stages w.r.t. the
post-fabricated device level in Figure \ref{fig:Comparison of the leakage assessment at pre-silicon stage}. In the pre-silicon stage, as the leakage assessment accuracy increases, the leakage evaluation time increases exponentially from RTL to gate-level (GTL) to layout level. It can also be observed that the flexibility in making design changes is reduced from RTL to gate-level to layout level. On the other hand, when observing the post-silicon stage, leakage assessment can be performed efficiently and accurately, however, flexibility for making design changes is very difficult (as in FPGA) if not impossible (as in ASIC).

\textbf{Our Contributions:} We propose a framework named RTL-PSC which can automatically assess PSC vulnerability at the earliest pre-silicon design stage, i.e., RTL. This framework is developed to be integrated into the traditional ASIC and FPGA design flow. RTL-PSC provides distinctive capabilities to chip designers and security analysts as listed below:

%First, they are developed for side-channel leakage assessment at post-silicon stage instead of pre-silicon stage. Since these techniques require large amount of plaintexts and power traces, a large simulation time would be required they are performed at pre-silicon stage. While it takes only a few seconds to apply thousands of plaintexts and collect the corresponding power traces at post-silicon, it may take days or months to perform similar simulation at the pre-silicon stage. Second, these techniques are developed for analyzing the whole design instead of localizing and identifying vulnerable blocks.

%Huss et al. \cite{DBLP:conf/birthday/HussSZ13}, \cite{jlpea7010004} proposed a side-channel vulnerability evaluator, named AMASIVE framework. AMASIVE identifies the hypothesis function for Hamming Weight/Hamming Distance model which is used for side-channel vulnerability assessment. However, this framework can only identify the hypothesis function, and the final vulnerability assessment still needs to be performed on a cryptographic hardware device. Also, designers have to be involved to identify the types of operation (e.g., permutation, substitution) each module performs. 

\begin{itemize}[leftmargin=*]
    \item   {\bf Leakage assessment at early design stage.}  The RTL-PSC framework is developed to automatically evaluate PSC vulnerability of a design at higher levels of abstraction to reduce time-to-market, cost of redesign, and the overall cost of adding security to the design.
    \item   {\bf Technology independent.} The vulnerability analysis is performed on RTL design to make the evaluation framework fairly library independent.
    \item   {\bf Fine granularity evaluation.} The RTL-PSC framework identifies which block of a design contributes the most to the vulnerability, i.e., pinpoints which block is leaking more secret information. Countermeasures only need to be applied for the vulnerable blocks/modules to reduce area overhead significantly.
    \item   {\bf Comprehensive analysis.} RTL-PSC considers the relation between blocks/modules of a design during vulnerability analysis instead of analyzing blocks/modules of a design independently.
    \item   {\bf Fast power estimation.} RTL-PSC estimates power leakage distribution based on the number of transitions at RTL to ensure the evaluation framework is fast.
    \item   {\bf Generic framework.} The RTL-PSC framework can be automatically applied to any cryptographic implementations at RTL without any customization or designers'  involvement.
    \item   {\bf Time, accuracy, and flexibility trade-off.} RTL-PSC can accurately and efficiently estimate PSC vulnerability at RTL. RTL-PSC has an average evaluation time of $35mins$ and its evaluation results closely matches with silicon results obtained from FPGA. Also, RTL-PSC provides the hardware designers with the most flexibility to address PSC vulnerabilities having the capability of working at RTL.
\end{itemize}

The rest of the paper is organized as follows: In Section II, the RTL-PSC framework is presented. Section III presents the RTL side-channel vulnerability evaluation metrics. Section IV provides and analyzes in detail the results demonstrating the performance of RTL-PSC and its respective metrics proposed in this paper. Finally, concluding remarks are offered in Section V. 

\section{RTL-PSC vulnerability evaluation framework}

\begin{figure}[!b]
  % \vspace{-3mm}
   \centering
   \includegraphics[width=0.45\textwidth]{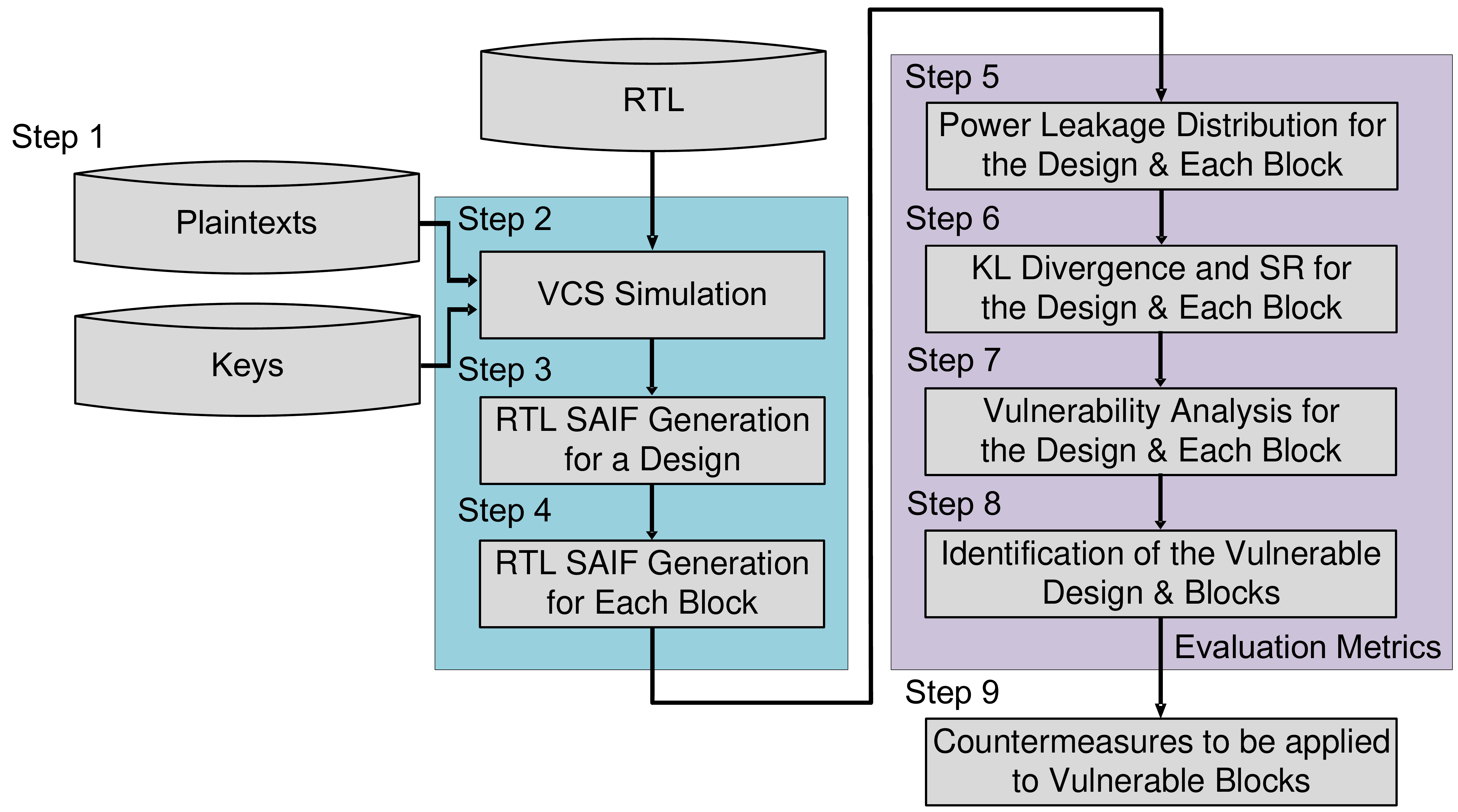}
   \caption{RTL-PSC framework.}
   \label{fig:RTL-PSC}
  % \vspace{-4mm}
\end{figure}

Figure \ref{fig:RTL-PSC} outlines the side-channel vulnerability evaluation framework. It includes two main parts, RTL Switching Activity Interchange Format (SAIF) file generation shown in the blue box and identification of vulnerable designs and blocks shown in the purple box. Algorithm \ref{alg:RTL_SC_vul} describes the identification technique for vulnerable designs and blocks. Note that, here $TC$ refers to the transition count, $\mathcal{N}$ refers to the Gaussian distribution, $f$ refers to the probability density function (PDF), $D_{KL}$ refers to the KL divergence and $ML$ refers to the maximum likelihood. Specifically, in Step 1, a group of simulation keys are specified. In Step 2, we utilize Synopsys VCS to perform functional simulation of the RTL design with the plaintexts and the selected keys as the inputs (key selection process is described in Section \ref{sec:key_pair}). In Step 3, once the simulation is complete, the SAIF file for the RTL design is generated. After finishing Step 3, all SAIF files for a group of keys and the applied plaintexts are generated (See Algorithm \ref{alg:RTL_SC_vul}, Lines 4-8). As its name indicates, the SAIF file includes the switching activity information for each net and register in the RTL design. Moreover, the SAIF file generated based on the RTL design has the same hierarchy as the design itself, hence, in Step 4, the SAIF file for each module in the design can be separated for localized vulnerability analysis. Next, the evaluation metrics are applied for leakage assessment. Specifically, in Step 5, the obtained switching activity is exploited to estimate the power leakage distribution for the design and each module within it (Lines 10-11 in Algorithm \ref{alg:RTL_SC_vul}). In Step 6, the Kullback-Leibler (KL) divergence \cite{kullback1951} and success rate (SR) based on power leakage distribution are calculated for the design and each block (Lines 12-15 in Algorithm \ref{alg:RTL_SC_vul}). In Step 7, vulnerability analysis is performed for the design and each block. In Step 8, the vulnerable design is identified based on the analysis performed in the previous step. Then the vulnerable blocks in the design that are leaking information the most are identified for further processing. Step 7 and Step 8 correspond to Lines 16-18 in Algorithm \ref{alg:RTL_SC_vul}. Following this, the framework enters into Step 9, where countermeasures only need to be applied to the vulnerable block(s). Note that Step 9 is outside the scope in this paper.

\begin{algorithm}[!t]
\caption{Identifying Vulnerable Blocks}
\label{alg:RTL_SC_vul}
\begin{algorithmic}[1]
\footnotesize
	\Procedure{Identifying vulnerable blocks}{}\\
	\textbf{Input:} RTL design, $\mathcal{P} = \{Plaintext\}, \{Key_0, Key_1\}$\\
  \textbf{Output:} $Set_{Vulnerable}$

	\For {$Key_j \in \{Key_0, Key_1 \}$}
	\For {$Plaintext_i \in \mathcal{P}$}
		\State $SAIF^{Key_j}_i \gets VCS(Plaintext_i, Key_j)$ 		
	\EndFor
	\EndFor
	
	\ForAll{$Block_i \in \mathcal{M}$}
		\State $TC^{0}_{Block_i} \gets \{ SAIF^{Key_0}_1, \ldots, SAIF^{Key_0}_n \}$
		\State $TC^{1}_{Block_i} \gets \{ SAIF^{Key_1}_1, \ldots, SAIF^{Key_1}_n \}$
		\State $f^0_i \gets \mathcal{N}(\mu_{TC^0_{block_i}}, \sigma^2_{TC^0_{block_i}})$
		\State $f^1_i \gets \mathcal{N}(\mu_{TC^1_{block_i}}, \sigma^2_{TC^1_{block_i}})$
		\State $KL_i \gets D_{KL}(f^0_i, f^1_i)$
		\State $SR_i \gets ML(f^0_i, f^1_i, n \;\; Plaintexts)$
		\If {$SR_i > SR_{threshold}$ or $KL_i > KL_{norm. th}$}
			\State $Set_{Vulnerable} \gets Block_i$
		\EndIf
	\EndFor
	\EndProcedure
\end{algorithmic}
\end{algorithm}
\vspace{-2mm}

\section{Evaluation metrics}

In order to reduce time-to-market and the overall cost of adding security to the design, the RTL power side-channel vulnerability evaluation metrics are proposed to perform a design-time evaluation of PSC vulnerability. \textit{To be specific, Kullback-Leibler (KL) divergence metric and success rate (SR) metric based on maximum likelihood estimation are developed and combined to evaluate vulnerability of a hardware implementation.} The first evaluation metric, i.e., KL divergence metric, estimates statistical distance between two different probability distributions, which is defined as follows \cite{kullback1951}:

\vspace{-2mm}
\begin{equation}
\label{eq:KL}
D_{KL}(k_i || k_j) = \int f_{T|k_i}(t) \log \frac{f_{T|k_i}(t)}{f_{T|k_j}(t)} dt
\end{equation}
\vspace{-2mm}

\noindent
where $f_{T|k_i}(t)$ and $f_{T|k_j}(t)$ are the probability density functions of the switching activity given keys $k_i$ and $k_j$, respectively.

For instance, if power leakage probability distributions based on two different keys are distinguishable, KL divergence between these two distributions is high, which provides indication on how vulnerable the implementation is. Hence, KL divergence is suitable for the vulnerability comparison between different implementations. However, the KL divergence value may be difficult to interpret when performing vulnerability analysis for just one implementation. To address this issue, we introduce the second evaluation metric, named success rate (SR)\footnote{This SR is the theoretical SR with infinite plaintexts and $\mathrm{SR}_{em}$ in Section \ref{sec:eval_metics} represents the empirical SR based on actual SCA attacks with $n$ plaintexts.} metric based on maximum likelihood estimation. The SR value represents the probability to reveal the correct key and is suitable to evaluate vulnerability for only one implementation. We derive the SR metric as follows: we assume that the probability density function of the switching activity $T$ given a key $K$ follows a Gaussian distribution, which can be expressed as follows:

\vspace{-2mm}
\begin{equation}
\label{eq:prob}
f_{T|K}(t)= \frac{1}{\sqrt{2 \pi} \sigma_k}e^{-\frac{(t - \mu_k )^2}{2\sigma_k^2}}
\end{equation}
%\vspace{-2mm}
where $\mu_k$ and $\sigma^2_k$ are the mean and variance of $T$, respectively.
The likelihood function is defined as $\mathcal{L}(k; t) = \frac{1}{n} \sum_{i=1}^n \ln f_{T|K}(t_i)$. Based on the maximum likelihood estimation, an adversary typically selects a guess key $\hat{k}$ as follows:

\vspace{-2mm}
\begin{equation}
\label{eq:ML}
\hat{k} = \operatorname*{arg\,max}_{k \in K} \mathcal{L}(k; t) = \operatorname*{arg\,max}_{k \in K} \frac{1}{n} \sum_{i=1}^n \ln f_{T|K}(t_i)
\end{equation}
\vspace{-2mm}

If the guess key ($k_g = \hat{k}$) is equal to the correct key ($k^*$), the side-channel attack is successful. Thus, the success rate can be defined as follows:

\vspace{-2mm}
\begin{equation}
\label{eq:sr1}
SR = \mathrm{Pr}[k_g = k^*] = \mathrm{Pr}[\mathcal{L}(k^*;t) > \mathcal{L}(\langle \bar{k^*} \rangle; t)]
\end{equation}
\vspace{-2mm}

\noindent
where $\langle \bar{k^*} \rangle$ denotes all wrong keys, i.e., the correct key $k^*$ is excluded from $\{k_1, k_2, \ldots, k_{n_k - 1}\}$.

KL divergence is closely related to SR since the mathematical expectation of $\mathcal{L}(k^*;t) - \mathcal{L}( k_i ; t)$ in Equation (\ref{eq:sr1}) is equal to KL divergence between $T|k^*$ and $T|k_i$ \cite{cryptoeprint:2014:152}. Hence, SR increases accordingly as KL divergence increases. Due to the relation between KL divergence and SR based on maximum likelihood estimation, the combination of KL and SR is proposed for leakage assessment. %However, SR can be used for the scenario with more than two keys, on the contrary, KL divergence can only be used for the scenario with exactly two keys.

\subsection{Selection of a Key Pair}
\label{sec:key_pair}

As shown in Algorithm \ref{alg:RTL_SC_vul}, first, a key pair is specified, then the probability distributions of the switching activity based on that key pair can be estimated using Equation (\ref{eq:prob}). The best key pair among all possible pairs is expected to provide the maximum KL divergence for vulnerability evaluation in the worst-case scenario. However, it is impossible and impractical to find the best key pair since the key space is huge, i.e., $\binom{2^{128}}{2}$. Alternatively, an appropriate key pair is able to be chosen, which satisfies the following conditions:

\begin{enumerate}
\item Assuming that each set of plaintexts is randomly generated, each key consists of the same subkey, e.g., $Key_0 = \{subkey \ldots subkey\} = 0x151515 \ldots 15$.
\item Hamming distance (HD) between two different subkeys is maximum, i.e., $subkey$ and $\overline{subkey}$.
\item If $D_{KL}(Key_0 || Key_i)$ increases asymptotically as $i$ increases, $i = 1,\ldots n$,  $Key_0$ and $Key_{n}$ are the appropriate key pair, where $Key_i$ is defined as $[subkey \ldots \underbrace{\overline{subkey} \;\; \overline{subkey}}_\text{i times}]$.
\end{enumerate}

These key pairs can be used for the post-silicon validation. While it is difficult to measure the isolated leakage of each block by a random key pair at the post-silicon stage, the leakage of each block can be measured \iffalse\footnote{In this paper, we did not calculate evaluation metrics of each block at post-silicon stage}\fi at the pre-silicon stage through applying the key pairs satisfying the above conditions. Moreover, the evaluation metrics would create the worst-case scenario through applying the key pairs with the maximum Hamming distance.

The selected keys applied to an AES RTL design are $16$ pairs of keys starting from all $0s$ key until all $Fs$ key. Each key has 128-bit and Hamming-distance between $Key_i$ and $Key_{i+1}$ is eight, which is shown in Table \ref{Simulation keys for evaluation framework}. Also, to take into account not only the key's impact on the power consumption, but also the plaintext's impact on power consumption, we use one thousand random plaintexts with the selected key pairs. We use the AES cipher operation itself for the generation of pseudo random plaintext \cite{keller2005nist}. We use $Plaintext_0$ as seed and then use each ciphertext ($j$) as the next plaintext $(j+1)$,\footnote{This seed is the same as TVLA's setup \cite{becker2013test} and 1000 plaintexts are enough to estimate SCA leakage based on our experiments.}

\vspace{-1mm}
\begin{equation} \label{eq:er1}
    \begin{aligned}
    & Plaintext_0 = 000 ... 000 \\
    & Plaintext_{j+1} = AES(Key_i, Plaintext_j), j = 0, 1, ... , 999.
    \end{aligned}
\end{equation}
\vspace{-2mm}

It can be noted that the key pair $Key_0$ and $Key_{16}$ in Table \ref{Simulation keys for evaluation framework} satisfies the above conditions. Furthermore, it can be seen that $Key_0$ would create a state similar as the reset state of the design, hence, $Key_0$ and $Key_{16}$ would create the worst case scenario. Therefore, this key pair is used for both the evaluation and the validation metrics, which is shown in Section \ref{sec:results}.

\begin{table}[!htbp]
    \vspace{-2mm}
    \centering
    \caption{Keys used in RTL-PSC framework.}
    \begin{tabular}
    {|c|c|}
    %{|>{\centering}m{1.65cm}<{\centering}|{m{1.85cm}<{\centering}|}
    \hline
    $Key_0$                        & 0x0000\_0000\_0000\_0000\_0000\_0000\_0000\_0000                      \\ \hline
    $Key_1$                        & 0x0000\_0000\_0000\_0000\_0000\_0000\_0000\_00FF                      \\ \hline
    %$Key_2$                        & 0x0000\_0000\_0000\_0000\_0000\_0000\_0000\_FFFF                      \\ \hline
    \multicolumn{2}{|c|}{......}                                                                         \\ \hline
    %$Key_{14}$                       & 0x0000\_FFFF\_FFFF\_FFFF\_FFFF\_FFFF\_FFFF\_FFFF                      \\ \hline
    \multicolumn{1}{|l|}{$Key_{15}$} & \multicolumn{1}{l|}{0x00FF\_FFFF\_FFFF\_FFFF\_FFFF\_FFFF\_FFFF\_FFFF} \\ \hline
    \multicolumn{1}{|l|}{$Key_{16}$} & \multicolumn{1}{l|}{0xFFFF\_FFFF\_FFFF\_FFFF\_FFFF\_FFFF\_FFFF\_FFFF} \\ \hline
    \end{tabular}
    \label{Simulation keys for evaluation framework}
    \vspace{-3mm}
\end{table}

\subsection{Identification of Vulnerable Designs and Blocks}

When the appropriate key pair is applied to a design, the same subkey patterns will be propagated to the same blocks, e.g., Sbox blocks in the AES design. The switching activity of each block and the entire design is recorded into SAIF files using VCS functional simulation, which corresponds to power leakage at RTL. The higher the difference between two power leakage distributions is, the higher impact the key has on the power consumption of the design/blocks, the more susceptible the design/blocks are to power analysis attack. Using the KL divergence and SR metrics, the vulnerable design and blocks within the design can be identified. If KL divergence or SR of any design or block is greater than $KL_{threshold}$ or $SR_{threshold}$, the design or the block is considered to be the vulnerable one (Line 16 in Algorithm \ref{alg:RTL_SC_vul}). $SR_{threshold}$ is determined based on the security constraint, e.g., $95 \%$ SR with $n$ plaintexts, while $KL_{threshold}$ is determined based on the $SR_{threshold}$ through the relation between KL divergence and SR.

\section{Results and analysis}
\label{sec:results}

In this section, we perform side-channel vulnerability assessment of two different implementations of AES algorithm using RTL-PSC. First, we provide a brief description of the two AES designs: AES Galois Field (GF) and AES Lookup Table (LUT). Then we present the evaluation results generated by RTL-PSC. We validate the accuracy of RTL-PSC evaluation results using gate-level simulation and FPGA silicon results.

\subsection{AES Benchmarks}

RTL-PSC framework is applied to AES-GF \cite{web:aes_gf} and AES-LUT \cite{web:aes_lut} encryption designs. Both are open-source designs. In AES-GF design, the AES key expansion and AES round operations occur in parallel. The AES-GF implementation takes 10 clock cycles to encrypt each data block. In contrast, the AES-LUT design first performs the key expansion and stores the expanded key in the key registers. The key expansion takes place once for each key. After the key expansion, the round operation starts and takes 11 clock cycles to encrypt a plaintext, precisely, one clock cycle for XORing a plaintext and the key, and 10 clock cycles for 10 round operations. Furthermore, the AES-GF architecture implements the AES SubByte operation with Galois-field arithmetic, while the AES-LUT architecture implements the AES SubByte operation with a lookup table.

The hierarchy of AES-GF and AES-LUT designs is as follows. The AES-GF consists of five SubByte blocks and four MixColumn blocks. Each SubByte block includes four Sbox blocks, each of which includes a GFinvComp block. On the other hand, the AES-LUT cipher consists of a SubWord block, a SubByte block and four MixColumn blocks. SubWord block performs the key expansion operation and therefore, is not related with the plaintext and it is not considered as a potential vulnerable block in the following analysis.

%\begin{figure}[!htbp]
%   \vspace{-2mm}
%   \centering
%   \includegraphics[width=0.45\textwidth]{./figures/figure6.pdf}
%   \caption{The hierarchy of the AES-GF and AES-LUT implementations.}
%   \label{fig:The hierarchy of the AES-GF and AES-LUT implementations}
%   \vspace{-2mm}
%\end{figure}

\subsection{Analyzing Suitability of the Key Pairs}
\label{sec:key_pair_result}

\begin{figure*}[!htbp]
\begin{center}
\begin{minipage}[c]{1\textwidth}
%\centering
%\begin{tabular}{@{}c@{}c@{}}
\begin{tabular}{m{0.47\textwidth}m{0.47\textwidth}}
%\includegraphics[width=0.5\textwidth,height=0.3\textwidth]{./figures/figure7a.eps}
%& \includegraphics[width=0.5\textwidth,height=0.3\textwidth]{./figures/figure7c.eps} \\
\includegraphics[width=0.45\textwidth]{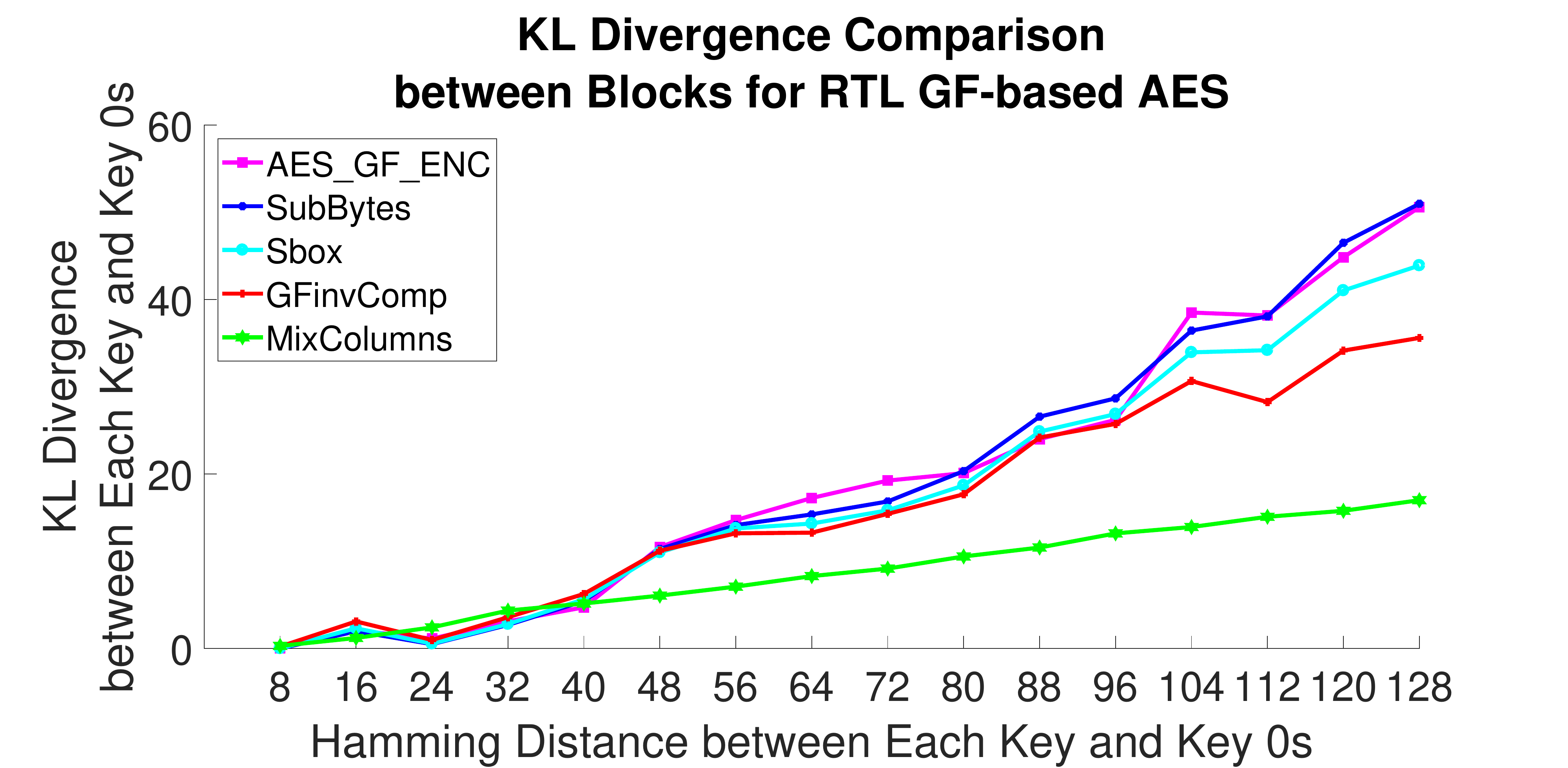}
%& \includegraphics[width=0.45\textwidth]{./figures/figure7c-eps-converted-to.pdf} \\
(a) \small KL divergence comparison between blocks for RTL AES-GF implementation % & (c) \small KL divergence comparison between blocks for GTL AES-GF \\
%\includegraphics[width=0.5\textwidth,height=0.3\textwidth]{./figures/figure7b.eps}
%& \includegraphics[width=0.5\textwidth,height=0.3\textwidth]{./figures/figure7d.eps} \\
&\includegraphics[width=0.45\textwidth]{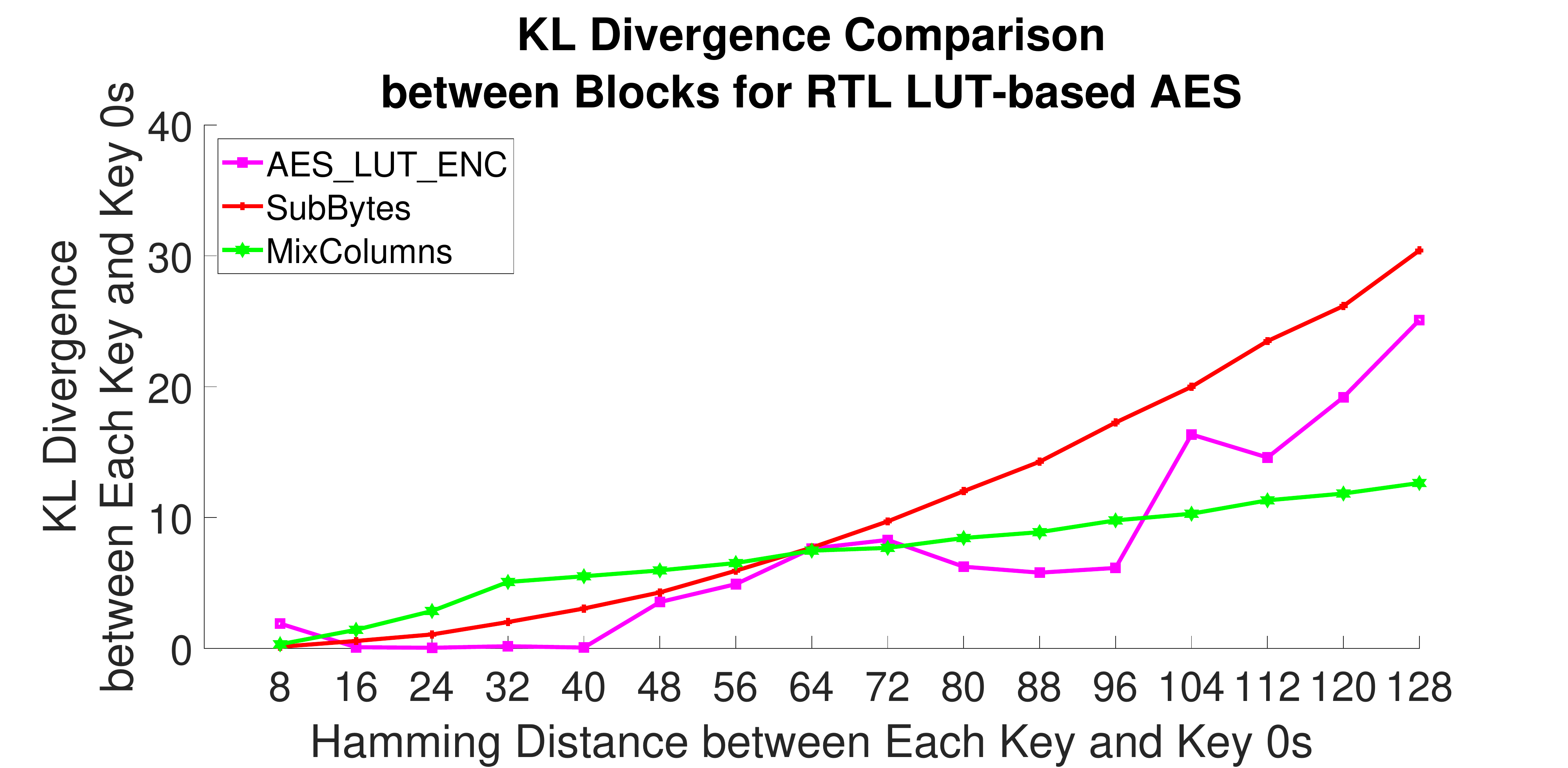}
%& \includegraphics[width=0.45\textwidth]{./figures/figure7d-eps-converted-to.pdf} \\
(b) \small KL divergence comparison between blocks for RTL AES-LUT implementation %& (d) \small KL divergence comparison between blocks for GTL AES-LUT \\
\end{tabular}
\caption {KL divergence comparison between blocks for RTL AES-GF and AES-LUT implementations.}
\label{fig:KL Divergence Comparison between Blocks for RTL/GTL AES-GF AND AES-LUT}
\end{minipage}%
\end{center}
%\vspace{-6mm}
\end{figure*}

We first analyze that the key pairs selected for RTL-PSC evaluation adheres the conditions described in Section \ref{sec:key_pair}.
Figures \ref{fig:KL Divergence Comparison between Blocks for RTL/GTL AES-GF AND AES-LUT}(a) and \ref{fig:KL Divergence Comparison between Blocks for RTL/GTL AES-GF AND AES-LUT}(b) illustrate the leakage assessment results at RTL. The switching activities of all blocks during 11 clock cycles are calculated. In the AES-GF implementation, the KL divergence of the design (i.e., AES\_GF\_ENC) and each block increases asymptotically with increasing the Hamming distance of the key pairs as shown in Table \ref{Simulation keys for evaluation framework}. Similarly, in the AES-LUT implementation, the KL divergence of the design (i.e., AES\_LUT\_ENC) and each block also increases asymptotically as the Hamming distance of the key pairs increases. It can be observed that the specified key pair ($key_0 = 0x00 \ldots 00, key_{16} = 0xFF \ldots FF$) satisfies the key selection conditions, hence is appropriate for RTL side-channel leakage assessment.

\subsection{RTL Evaluation Metrics}
\label{sec:eval_metics}

\begin{figure*}[h]
\begin{center}
\begin{minipage}[c]{1\textwidth}
\begin{tabular}{m{0.48\textwidth}m{0.48\textwidth}}
\includegraphics[width=0.4\textwidth,height=0.2\textwidth]{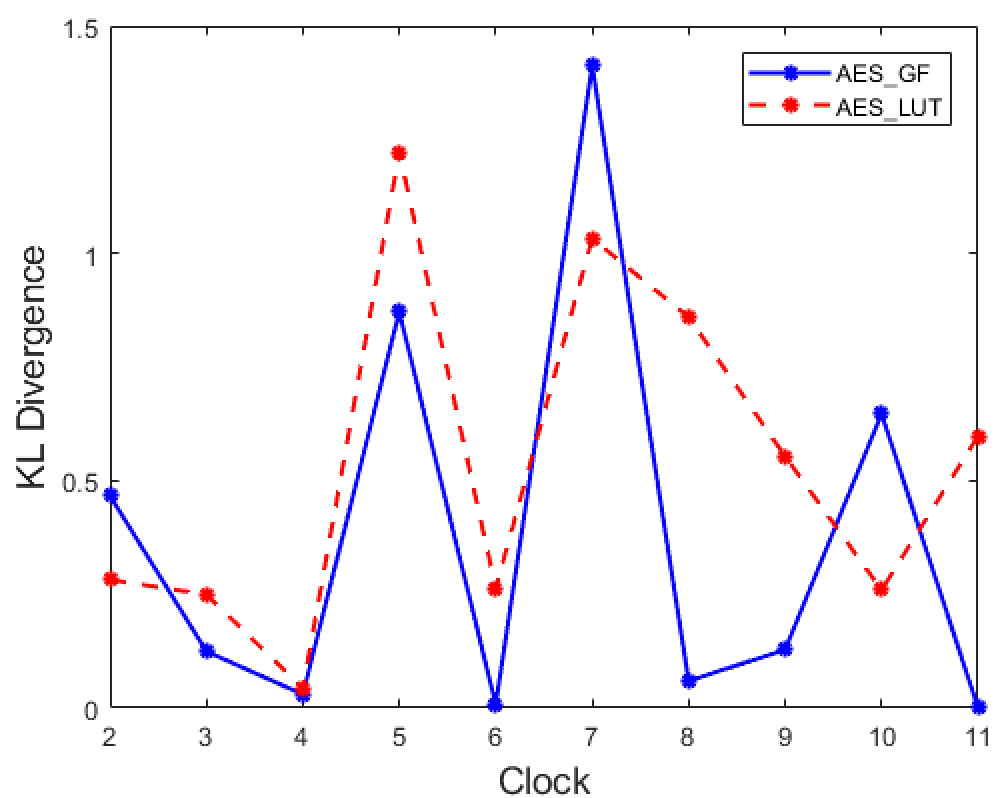}
& \includegraphics[width=0.4\textwidth,height=0.2\textwidth]{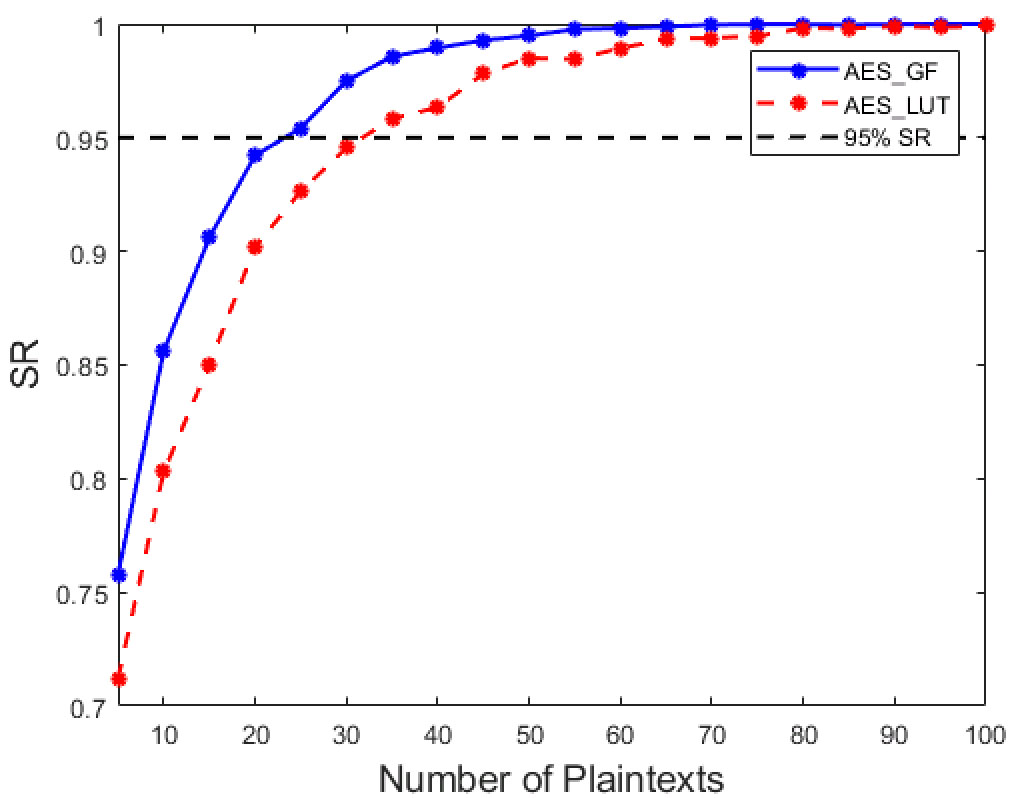} \\
(a) \small KL divergence per clock cycle for AES-GF and AES-LUT implementations & (b) \small $\mathrm{SR_{em}}$ corresponding to KL divergence (0.47 and 0.28) for AES-GF and AES-LUT implementations\\
\end{tabular}
\caption {KL divergence and SRs for AES-GF and AES-LUT implementations.}
\label{fig:KL_SR_GF_LUT}
\end{minipage}%
\end{center}
%\vspace{-6mm}
\end{figure*}

\begin{figure*}[h]
\begin{center}
\begin{minipage}[c]{1\textwidth}
\begin{tabular}{m{0.55\textwidth}m{0.4\textwidth}}
\includegraphics[width=0.55\textwidth,height=0.3\textwidth]{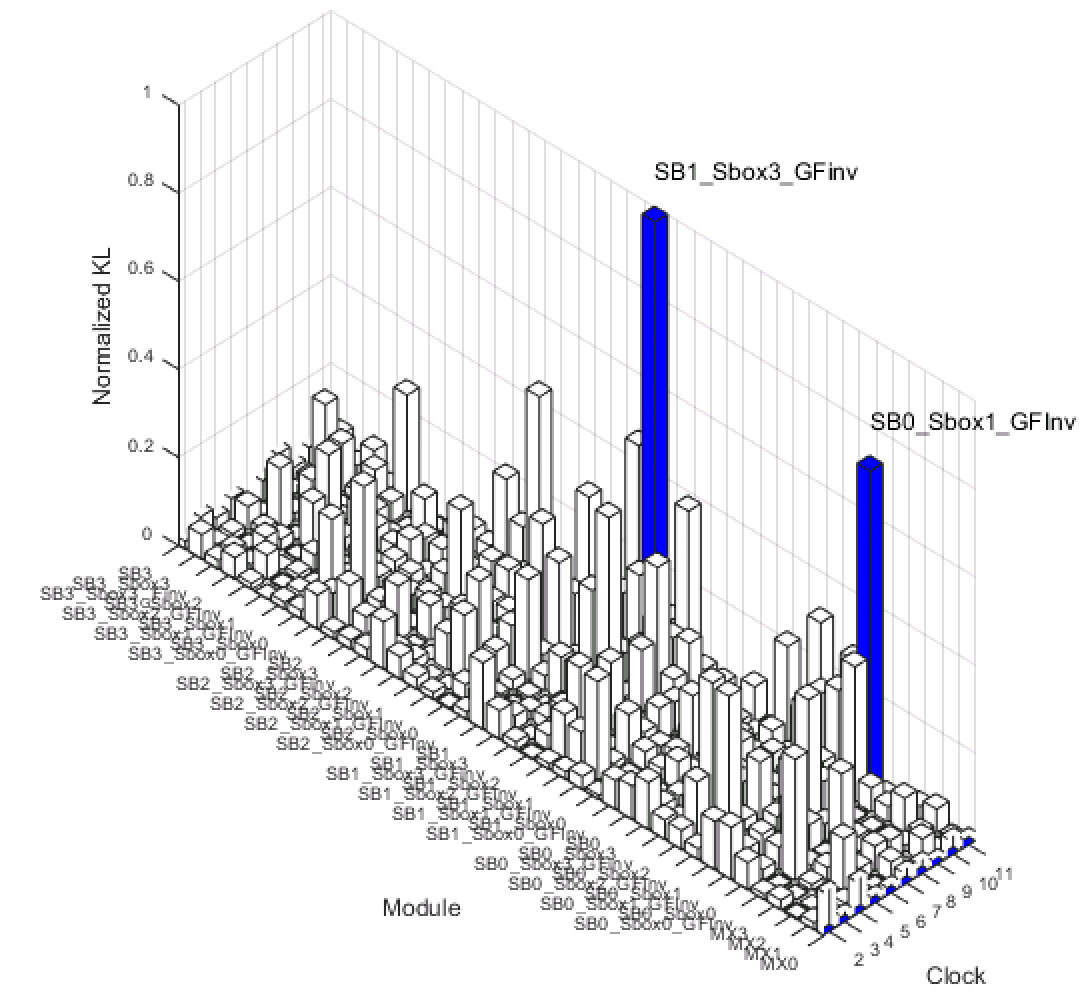}
& \includegraphics[width=0.4\textwidth,height=0.3\textwidth]{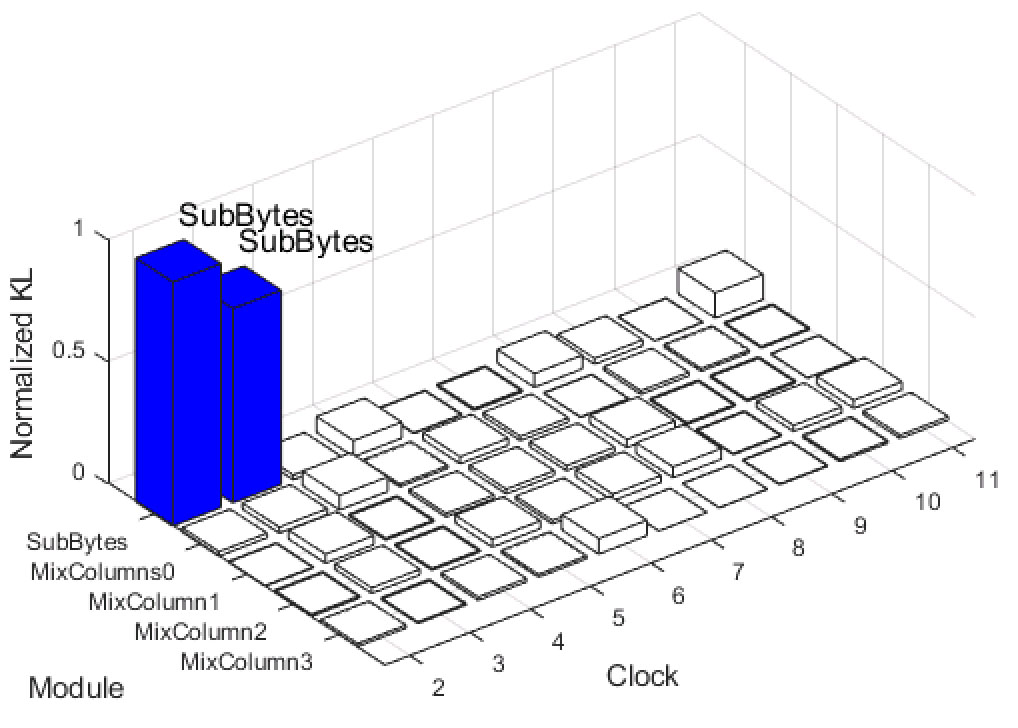} \\
(a) \small Normalized KL divergence for AES-GF implementation in both time and spatial/modular domains & (b) \small Normalized KL divergence for AES-LUT implementation in both time and spatial/modular domains \\
\end{tabular}
\caption {Normalized KL divergence for vulnerable blocks within AES-GF and AES-LUT implementations ($KL_{norm. th} = 0.5$).}
\label{fig:KL_xy_GF_LUT}
\end{minipage}%
\end{center}
%\vspace{-8mm}
\end{figure*}

In order to determine if a RTL design is vulnerable, KL divergence of the design per clock cycle is calculated. Figure \ref{fig:KL_SR_GF_LUT}(a) shows KL divergence from the second clock cycle to the 11th clock cycle of both AES-GF and AES-LUT RTL implementations, during which 10 round operations for an encryption are performed.  At the second clock cycle corresponding to the first round operation, which is mostly exploited by power analysis attack, KL divergence of AES-GF and AES-LUT implementations are 0.47 and 0.28, respectively. These values correspond to $95 \%$ $\mathrm{SR}_{em}$\footnote{The $\mathrm{SR}_{em}$ represents the empirical SR based on actual SCA attacks with $n$ plaintexts.} vulnerability level ($SR_{threshold}$) with $25 \; plaintexts$ and $35 \; plaintexts$, respectively, as shown in Figure \ref{fig:KL_SR_GF_LUT}(b). Based on KL divergence metric, as shown in Figure \ref{fig:KL_SR_GF_LUT}(a), KL divergence of AES-GF implementation at the second clock cycle (i.e., 0.47) is greater than that of AES-LUT implementation (i.e., 0.28). Likewise, based on SR metric, as shown in Figure \ref{fig:KL_SR_GF_LUT}(b), the DPA attack success rate of AES-GF implementation is higher than that of AES-LUT implementation with the same number of plaintexts. Similarly, the number of plaintexts required for successful DPA attack on AES-GF implementation is less than that on AES-LUT implementation. Hence, compared with AES-LUT implementation, AES-GF implementation is identified as the more vulnerable design. %{\color{red} Furthermore, based on the KL divergence and SR evaluation metrics, if the DPA attack success rate for a secure implementation is defined as $95\%$, the KL divergence is required to be less than 0.47 ($KL_{threshold}$) to ensure the success rate is less than $95\%$, thus achieving a secure AES-GF implementation.}

\textbf{Vulnerable Block Identification:} Once a RTL deign is determined as a vulnerable one, the next step is to identify the vulnerable blocks within the design that contribute to side-channel leakage significantly. Side-channel vulnerability can be evaluated in both time and spatial/modular domains. In other words, the evaluation metrics based on the switching activity of each block within the design are calculated at fine-granularity scale so that vulnerable blocks can be identified per clock cycle.

First, KL divergence in both time and spatial/modular domains is normalized, i.e., KL divergence of each block is divided by the maximum KL divergence of those blocks ($KL_{norm. th} = KL_i / max(KL_i)$). Then, if the normalized KL divergence of any block is greater than $KL_{norm. th} = 0.5$, that block is identified as the vulnerable one and included into the set of vulnerable blocks. Figure \ref{fig:KL_xy_GF_LUT} shows KL divergence of each block in both time and spatial/modular domains. The identified vulnerable blocks (i.e., KL divergence greater than $KL_{norm. th} = 0.5$) are denoted with blue bars. Specifically, in the AES-GF design, GFinvComp blocks within Sbox0 and Sbox1 blocks are identified as the vulnerable ones; in the AES-LUT design, SubByte blocks are identified as the vulnerable ones. It should be noted that the threshold values ($KL_{threshold}$ and $KL_{norm. th}$) can be adjusted by the SR vulnerability level.

\textbf{Evaluation Time:} The evaluation time of RTL-PSC includes VCS functional simulation time of the RTL design as well as the data processing time required for analyzing the SAIF files. The evaluation time of RTL-PSC for AES-GF is 46.3 minutes and for AES-LUT is 24.03 minutes. If the same experiments were performed at gate-level designs, it would take around 31 hours. Therefore, our RTL-PSC is almost 42X more efficient as compared to similar gate-level assessment. The evaluation time at layout level is going to be even more expensive (more than a month). RTL-PSC also provides the flexibility to make design changes whereas, the post-silicon assessment provides no flexibility. In the next subsection, we will validate that RTL-PSC is not only efficient, but also accurate using the post-silicon results.

\begin{table*}[t]
    %\vspace{-3mm}
    \centering
    \caption{Correlation coefficient between KL divergence of RTL and GTL blocks.}
    \begin{tabular}{|c|c|c|c|c|c|}
    \hline
    \multicolumn{4}{|c|}{AES-GF Blocks, RTL vs. GTL} & \multicolumn{2}{c|}{AES-LUT Blocks, RTL vs. GTL} \\ \hline
    SubByte    & Sbox       & GFinvComp    & MixColumn   & SubByte                  & MixColumn                 \\ \hline
    99.11\%     & 99.55\%    & 99.64\%      & 94.73\%      & 99.71\%                   & 96.80\%                    \\ \hline
    \end{tabular}
    \label{Correlation coefficient between KL divergence at RTL and at GTL for AES-GF and AES-LUT at block level}
    %\vspace{-1mm}
\end{table*}

\subsection{Validation of RTL-PSC}
\label{sec:val}

The RTL-PSC results are validated through both gate-level and FPGA implementations. We present that the PSC vulnerability assessment results generated by RTL-PSC is highly correlated to the assessment results retrieved from gate-level and FPGA. 

\textbf{GTL Validation:} For gate-level validation, we first synthesize the RTL codes of AES-GF and AES-LUT to gate-level netlist using Synopsys Design Compiler \cite{synopsys} with Synopsys standard cell library. Next we utilize VCS to perform functional simulation of the netlist with the same plaintexts and keys as used in RTL, and generate the corresponding SAIF files. Then, we use Synopsys PrimeTime \cite{synopsys} as well as the generated SAIF files to report the power consumption for the entire design and each block inside the design. Finally, we derive the gate-level KL divergence metric for the design and each block using Equation \ref{eq:KL}. 

We then calculate the Pearson correlation coefficient between the KL divergence of the RTL and GTL design/blocks (shown in Column 2 of Table \ref{Correlation coefficient between KL divergence at RTL and at GTL and at FPGA silicon level for AES-GF and AES-LUT at design level} and Table \ref{Correlation coefficient between KL divergence at RTL and at GTL for AES-GF and AES-LUT at block level}). The high correlation coefficient values ($> 90\%$) indicate that the PSC vulnerability assessment results generated by RTL-PSC at RTL is almost as accurate as the assessment results retrieved from GTL. 

Next we validate the vulnerable block identification results produced by RTL-PSC. Table \ref{Correlation coefficient between KL divergence at RTL and at GTL for AES-GF and AES-LUT at block level} presents the correlation coefficient values between the KL divergence metric for each block at RTL and gate-level. The KL metric for each block of AES-GF and AES-LUT implementations at RTL has a high correlation to that at gate-level indicating that our vulnerable block identification technique at RTL is as accurate as gate-level. Next we validate the RTL-PSC evaluation results w.r.t. FPGA.

\begin{table}
    %\vspace{-3mm}
    \centering
    \caption{Correlation coefficient between KL divergence at RTL, GTL, and FPGA silicon level.}
    \begin{tabular}{|c|c|c|}
    \hline
    Benchmark & RTL vs. GTL & RTL vs. FPGA Silicon Level  \\ \hline
    AES-GF  &  99.57\%     & 98.83\%     \\ \hline
    AES-LUT & 90.35\%     & 80.80\%        \\ \hline
    \end{tabular}
    \label{Correlation coefficient between KL divergence at RTL and at GTL and at FPGA silicon level for AES-GF and AES-LUT at design level}
    %\vspace{-1mm}
\end{table}

\begin{figure}[b]
\centering
\includegraphics[width=.35\textwidth]{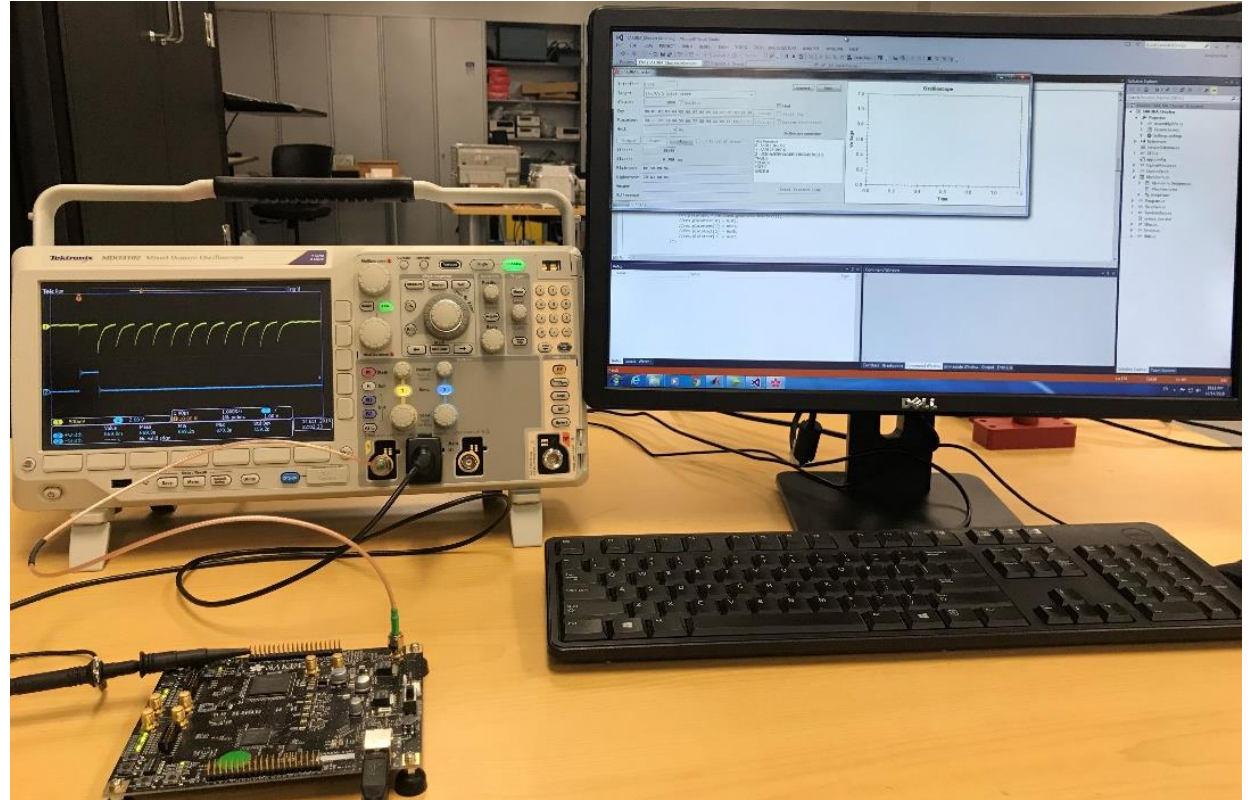}
\caption {Experimental setup for FPGA validation.}
\label{fig:setup}
\end{figure}

\textbf{FPGA Validation:} For FPGA silicon validation, we use the SAKURA-G board \cite{web:sakura_g} for AES implementations, which contains two SPARTAN-6 FPGAs and is designed for research and development on hardware security. Tektronix MDO3102 oscilloscope is used to measure the voltage drop between shunt registers connected to the Vdd pin. The clock frequency of the AES implementation is 24 MHz. The sampling rate and bandwidth of the oscilloscope are 500 MS/s and 250 MHz, respectively. Figure \ref{fig:setup} shows the experimental setup for the FPGA validation.

We first map the AES-GF and AES-LUT designs on an FPGA and then, apply the same plaintexts and keys as used at RTL, and measure the power consumption during encryption operation. Following this, we derive the KL divergence metric from the collected power traces. Then, we can calculate the Pearson correlation coefficient between the KL divergence at RTL and FPGA (as shown in Column 3 of Table \ref{Correlation coefficient between KL divergence at RTL and at GTL and at FPGA silicon level for AES-GF and AES-LUT at design level}). The high correlation coefficient values ($> 80\%$) indicate that the PSC vulnerability assessment results generated by RTL-PSC at RTL is almost as accurate as FPGA assessment. In other words, RTL-PSC can accurately analyze PSC vulnerability at RTL. 

Note that many implementation details, e.g., glitches caused by the gate delay, clock gating, datapath gating, retiming, clock and power network structure is not available at RTL unlike gate-level and FPGA implementations. In spite of these limitations, RTL-PSC is able to accurately identify PSC vulnerability which proves the efficacy of the framework.

Also, note that it is not feasible to perform vulnerable block identification at FPGA. The reason is that, it is not feasible to isolate the power traces associated to each block from the post-silicon power measurements. Therefore, we could not validate vulnerable block identification at FPGA.

\textbf{Comparison with State-of-the-art:} Veshchikov \textit{et al.} \cite{veshchikov2017use} presented a comprehensive survey of simulators for side-channel analysis, e.g., PINPAS \cite{den2003pinpas}, SCARD \cite{aigner2006side}, OSCAR \cite{thuillet2009smart}, etc. These simulators mostly support software crypto algorithms implemented in microprocessors, not hardware crypto modules. On the other hand, TVLA \cite{becker2013test} and $\chi^2$-test \cite{moradi2018leakage} can work with hardware crypto designs. However, these techniques only provide a pass/fail test and do not provide the quantitative amount of leakage. AMASIVE framework \cite{DBLP:conf/birthday/HussSZ13} identifies the hypothesis function for HW/HD model to be used for side-channel vulnerability assessment. The major limitation is that it can only identify the hypothesis function and the final vulnerability assessment still needs to be carried out on a prototype device. In contrast, RTL-PSC can quantitatively and accurately assess PSC leakage of hardware crypto modules in RTL level in time efficient manner.

\section{Conclusion and Future Work}
In this paper, we proposed an automatic evaluation framework to perform a design-time evaluation of side-channel attacks resistance for a cryptographic implementation at RTL. Instead of measuring dynamic power, switching activity at RTL is exploited by using VCS functional simulation to estimate power profile to ensure the evaluation framework is efficient, effective, and library independent. Once the power estimation is complete, the KL divergence metric and SR metric based on maximum likelihood estimation are combined to identify the vulnerable design and blocks within the design. Experimental results on AES-GF and AES-LUT implementations demonstrated that the methodology proposed in this paper were able to perform leakage assessment and identify the vulnerable design/blocks efficiently, effectively, and precisely.

In our future work, the proposed evaluation framework will be applied to evaluate vulnerabilities of the DPA protected designs, thus providing information in terms of scaling and effectiveness for leakage assessment of protected ones. RTL-PSC will be an important component of the recent academic and industrial initiatives for developing CAD frameworks which aim at automating the security vulnerability assessment of hardware designs at design stages \cite{xiao2016security, nahiyan2017security, nahiyan2016avfsm, nahiyan2017hardware, nahiyan2018security}.

%\vspace{-1mm}
\footnotesize

\bibliographystyle{IEEEtran}
\bibliographystyle{plain}
\bibliography{references}

\end{document}